\documentclass{article}
\usepackage[utf8]{inputenc}

\usepackage{fullpage}
\usepackage{amsmath}
\usepackage{amssymb}
\usepackage{amsfonts}
\usepackage{amsthm}
\usepackage{enumitem}
\usepackage{graphicx}
\usepackage{todonotes}
\usepackage{xcolor}
\usepackage{hyperref}
\usepackage{caption}
\usepackage[labelfont=bf]{caption}
\usepackage[capitalise]{cleveref}
\usepackage{comment}
\usepackage{cases}
\usepackage{xfrac}

\renewcommand{\d}{\ensuremath{\mathrm{d}}}
\renewcommand{\tilde}{\widetilde}
\newcommand{\oh}{\sfrac{1}{2}}

\newcommand{\calF}{\mathcal{F}}
\newcommand{\calG}{\mathcal{G}}
\newcommand\norm[1]{\left\lVert {#1} \right\rVert}

\newcommand{\mybf}{}

\title{Application of Machine Learning and Convex Limiting to Subgrid Flux Modeling in the Shallow-Water Equations}
\author{
Ilya Timofeyev\thanks{Dept. of Mathematics, University of Houston, Houston, TX 77204, itimofey@cougarnet.uh.edu} \quad
  Alexey Schwarzmann \thanks{Dept. of Mathematics, TU Dortmund, alexey.schwarzmann@math.tu-dortmund.de} \quad
  Dmitri Kuzmin \thanks{Dept. of Mathematics, TU Dortmund,
kuzmin@math.uni-dortmund.de}
}
\date{}

\begin{document}

\maketitle

\begin{abstract}
  We propose a combination of machine learning and flux limiting for property-preserving subgrid scale modeling in the context of flux-limited finite volume methods for the one-dimensional shallow-water equations. The numerical fluxes of a conservative target scheme are fitted to the coarse-mesh averages of a monotone fine-grid discretization using a neural network to parametrize the subgrid scale components. To ensure positivity preservation and the validity of local maximum principles, we use a flux limiter that constrains the intermediate states of an equivalent fluctuation form to stay in a convex admissible set. The results of our numerical studies confirm that the proposed combination of machine learning with monolithic convex limiting produces meaningful closures even in scenarios for which the network was not trained.
  \medskip

\end{abstract}

\section{Introduction}
Many fluid flow models are based on nonlinear conservation laws that
incorporate a vast range of spatial and temporal scales.
Therefore, direct numerical simulations are often prohibitively expensive and,
despite rapid progress in computer architecture, existing codes cannot be run at the discretization level required to resolve all relevant physical processes. Therefore, the derivation/calibration of reduced models has been an active area of research for many decades. The simplest Large-Eddy Simulation (LES) approaches use the Smagorinsky eddy viscosity model \cite{smag63} for subgrid stresses that appear in the spatially filtered equations. More advanced alternatives include stochastic (e.g. \cite{leith96,mtv2,palmer2009backscatter,jansen2014,dat12,zadoacti18,berner2017,ress2022}) and deterministic (e.g. \cite{acbr99,germano91dsgs,lilly92dsgs,moin91dsgs}) closures, alpha models \cite{holm2003} etc. Recently, machine learning has also been used to represent the effects of unresolved degrees of freedom 
(e.g. \cite{boza19,alcala21,zanna2023,berloff2021}, review \cite{cfdml2023review}).

The main focus of our work is on the Shallow-Water Equations (SWE) \cite{saint1871theorie}. This nonlinear hyperbolic system is widely used to simulate flow in rivers and channels, as well as coastal and ocean dynamics. In the context of ocean dynamics, parametrization of mesoscale eddies requires developing appropriate subgrid scale models. Recent years have witnessed the advent of closures based on Machine Learning (ML). The validity and effectiveness of such modern closures was demonstrated by many numerical examples. However, the use of Neural Network (NN) approximations makes it more difficult to prove the stability, consistency, and convergence of ML numerical methods for nonlinear PDEs. Moreover, for the SWE model, an NN-based closure may violate entropy inequalities or produce negative water heights. As a fail-safe remedy, we consider an NN model for representing subgrid fluxes (instead of estimating the whole right-hand side of the reduced model) and propose the application of a physics-aware flux limiter to ML-generated subgrid fluxes that may violate important inequality constraints. 
We use a relatively simple feed-forward network to extract subgrid fluxes from data generated by the fully resolved model.
The monolithic convex limiting (MCL) strategy \cite{kuzmin2020monolithic} that we use in this work represents the coarse-mesh cell averages as convex combinations of intermediate states that belong to a convex admissible set. The proofs of desired properties exploit this fact following the analysis of flux-corrected transport (FCT) algorithms by Guermond et al. \cite{guermond2018second}.

The rest of the paper is organized as follows. We define the conserved variables and fluxes of the one-dimensional SWE system in Section \ref{sec:swe}. The space discretization of this nonlinear hyperbolic system and the definition of coarse-scale variables are discussed in Section \ref{sec:coarse}. The fine-mesh discretization uses the Local Lax--Friedrichs (LLF) flux and is bound preserving. In Section \ref{sec:nn}, we  construct the subgrid fluxes of the coarse-mesh model using a neural network. The MCL framework that we present in Section \ref{sec:mcl} is used for flux limiting purposes. In Section \ref{sec:num}, we perform numerical experiments. The results of fully resolved simulations are compared with NN-based reduced model approximations. The effect of locally bound-preserving MCL corrections is studied by switching the flux limiter on and off. Conclusions are drawn in Section \ref{sec:conc}.

\section{Shallow-Water Equations}
\label{sec:swe}
We consider the one-dimensional SWE system with periodic boundary conditions imposed on the boundaries of the domain $\Omega=(0,L)$. The evolution of the fluid depth $h(x,t)$ and fluid velocity $v(x,t)$ is governed by
\begin{equation}\label{eq:swe}
  \begin{bmatrix} h\\ hv \end{bmatrix}_t + \begin{bmatrix} hv \\
hv^{2}+\frac{1}{2}gh^{2} \end{bmatrix}_x = 0,
\end{equation}
where $x\in\bar\Omega$ is the space variable, $t\ge 0$ is the 
time variable, and $g$ is the gravitational acceleration.
The 
initial conditions are given by
$h(x,0)=h_0(x)$, $v(x,0)=v_0(x).$
We assume that $h(x,t) > 0$, i.e., no dry
states are present. 
Introducing the discharge $q = hv$,
the hyperbolic system \eqref{eq:swe} can be rewritten as 
\begin{equation}\label{eq:swe_conservative_form}
  \begin{bmatrix} h\\q \end{bmatrix}_t 
  + \begin{bmatrix} q\\ q^{2}/h+\frac{1}{2}gh^{2} \end{bmatrix}_x=0.
\end{equation}
This is a system of the form $u_t+ (f(u))_x = 0$, where $u=[h, q]^T$
is the vector of conserved quantities and $f(u)=[q, q^{2}/h+\frac{1}{2} gh^{2}]^T$ is the flux function. We initialize $u$ by $u_0 = [h_0, q_0]^T$, where $q_0 = h_0v_0$.

\section{Discretization, Coarse-Scale Variables, and Subgrid Fluxes}
\label{sec:coarse}
A finite-volume discretization of the SWE system
on a uniform fine mesh with spacing $\Delta x = L/N_f$
produces a system of ordinary differential equations
\begin{equation}\label{eq:semi-discrete-fv}
  \frac{\d}{\d t} u_{i}= \frac{F_{i-\oh}-F_{i+\oh}}{\Delta x}
  \qquad i=1, \ldots, N_f
\end{equation}
for approximate fine-mesh cell averages $u_i(t)\approx\frac{1}{\Delta x}
\int_{x_{i-\oh}}^{x_{i+\oh}}u(x,t)\mathrm{d} x$. The numerical flux
\begin{equation}\label{eq:llf-flux}
  F_{i+\oh}= \frac{f(u_{i})+f(u_{i+1})}{2}- \frac{\lambda_{i+\oh}}{2}(u_{i+1}-u_{i})
\end{equation}
of the Local Lax-Friedrichs (LLF) scheme is defined using the maximum
wave speed $\lambda_{i+\oh}$ of the Riemann problem with the initial
states $u_i$ and $u_{i+1}$. The intermediate state
\begin{equation}\label{eq:bar_state}
  \bar{u}_{i+\oh}:=\frac{u_{i+1}+u_i}{2}-\frac{1}{2\lambda_{i+\oh}}(f(u_{i+1})-f(u_i))
\end{equation}
represents a spatially averaged exact solution of such a Riemann problem.
In what follows, the state $\bar{u}_{i+\oh}$ is referred to
as a \emph{bar state}.
Using the maximum eigenvalue of the Jacobian matrices $f'(u_i)$ and
$f'(u_{i+1})$ as an approximate upper bound for the local wave
speed, we set \cite{leveque2002finite}
\begin{equation}\label{eq:wave_speed}
  \lambda_{i+\oh}:=\max \left( \lvert v_i\rvert+\sqrt{gh_i},\, \lvert
  v_{i+1}\rvert +\sqrt{gh_{i+1}} \right) .
\end{equation}

We denote by $\calF(u_L,u_R)$ the numerical flux function
such that $F_{i+\oh} = \calF(u_{i}, u_{i+1})$ is given
by \eqref{eq:llf-flux} with $\lambda_{i+\oh}$ defined by
 \eqref{eq:wave_speed}.
Substituting the so-defined LLF flux 
into \eqref{eq:semi-discrete-fv} and invoking the
definition \eqref{eq:bar_state} of $\bar{u}_{i+\oh}$, we write the spatial
semi-discretization \eqref{eq:semi-discrete-fv}
in the equivalent \emph{fluctuation form} (cf.  \cite{leveque2002finite})
\begin{equation}\label{eq:semi-discrete-llf}
  \frac{\d}{\d t} u_{i}=\frac{1}{\Delta x}\bigl[\lambda_{i-\oh}(\bar{u}_{i-\oh}-u_i)
    +\lambda_{i+\oh}(\bar{u}_{i+\oh}-u_i) \bigr],\qquad i=1,\ldots,N_f.
\end{equation}

In this paper, 
we utilize Heun's time-stepping method. This explicit two-step
Runge-Kutta scheme is second-order accurate and
yields a convex combination of two forward Euler predictors. 
Heun's method is a representative of \emph{strong stability preserving} (SSP)
methods \cite{shu1988efficient,gottlieb2001strong,gottlieb2011strong}.
The resulting space-time discretization also belongs to the class of 
\emph{height positivity preserving} (HPP) methods, i.e.,
the fully discrete scheme guarantees that the water height $h_i(t)$ remains strictly 
positive for all times (see \cite{alexey_thesis} for details).

 We assume that the fine-mesh discretization resolves all physical processes of interest sufficiently well.
 This, essentially, implies that the number $N_f$ of fine mesh cells
\[
C_i=[x_{i-\oh}, x_{i+\oh}], \quad i=1,\ldots,N_f
\]
is very large and the
 numerical viscosity introduced by the LLF scheme is very small. 
To avoid the high cost of direct numerical simulations on the fine mesh,
we compose a coarse mesh from macrocells
\begin{equation}
  \bar{C}_{i}:=\bigcup_{j=k(i-1)+1}^{ki}C_{j} = [x_{k(i-1)+\oh},\,x_{ki+\oh}]
  , \qquad i=1,\ldots, N_c,
\end{equation}
where $k$ is the coarsening degree, $N_c=N_f/k$, and $\Delta X=k\Delta x$.
The coarse-scale variables 
\begin{equation}
  \label{eq:resolved_mode}
  U_{i}(t) = \frac{1}{k}\sum_{j=k(i-1)+1}^{ki}u_{j}(t), \qquad i=1,\ldots, N_c
\end{equation}
correspond to averages over the macrocells $\bar{C}_{i}$. In principle, we could use
the LLF finite volume discretization of the integral
conservation law for $\bar{C}_{i}$
to directly obtain an evolution equation for $U_{i}$. However,
a coarse-mesh LLF approximation is unlikely to resolve all physical processes appropriately even for moderate values of the parameter $k$. To derive \emph{subgrid fluxes} that significantly reduce the error of the
coarse-grid LLF scheme, we need to average equations
\eqref{eq:semi-discrete-llf}. The resulting
equation for $U_i$ can be written in the form
\begin{equation}\label{eq:evolUi}
  \frac{\d }{\d t}U_i =\frac{(\bar{F}_{i-\oh}+G_{i-\oh})
  -(\bar{F}_{i+\oh}+G_{i+\oh})}{\Delta X},
\end{equation}
where  $\bar{F}_{i+\oh} = \calF(U_{i}, U_{i+1})$ is
 the coarse-mesh LLF flux. The fine-scale component
of the subgrid flux
\begin{equation}
  \label{eq:subgrid_flux}
  G_{i+\oh}:=F_{ki+\oh}-\bar{F}_{i+\oh}
\end{equation}
is given by
$F_{ki+\oh}=\calF(u_{ki},u_{ki+1})$. The point $X_{i+1/2}=x_{ki+\oh}$ 
represents the interface
of fine-mesh cells $C_{ki}$ and $C_{ki+1}$ such that 
$C_{ki}\cap C_{ki+1}=\bar{C}_i\cap\bar{C}_{i+1}$. The flux defined by
\eqref{eq:subgrid_flux} can be also expressed as
\begin{equation}
  \label{eq:subgrid_flux_function}
  G_{i+\oh}:=
  \calG(u_{ki}, u_{ki+1}, U_{i}, U_{i+1}) = 
  \calF(u_{ki},u_{ki+1})
  -\calF(U_{i}, U_{i+1}).
\end{equation}
In general, 
the subgrid flux depends on the coarse-scale variables and the fine-mesh cell averages.
The evolution equation \eqref{eq:evolUi} has the structure of a finite-volume discretization of the SWE system on a coarse mesh. The coarse-mesh LLF discretization without the subgrid fluxes is given by
\[
  \frac{\d }{\d t}U_i =\frac{\bar{F}_{i-\oh} 
  -\bar{F}_{i+\oh}}{\Delta X}.
  \]
  Since it is usually more diffusive than the fine-mesh discretization
  \cite{leveque2002finite}, the subgrid fluxes $G_{i+\oh}$
  can be interpreted as antidiffusive corrections to the
  low-order flux approximations $\bar{F}_{i+\oh}$.

To obtain a closed-form equation for the coarse-scale variables, 
we need to construct a suitable approximation to the subgrid flux
in terms of the coarse-mesh data. 
It is reasonable to expect that the subgrid flux $G_{i+\oh}$ depends on the
coarse-mesh cell averages $U_i$ and $U_{i+1}$ but, in our experience, two-point
stencils do not provide enough data for subgrid flux modeling using
neural networks. Therefore, following Alcala \cite{alcala2021subgrid_phd},
we use a larger stencil $[U_{i-1}, U_i, U_{i+1}, U_{i+2}]$ for the
computation of the subgrid flux $G_{i+\oh}$. 
The coarse-mesh variables are used as input to the neural network
that generates an approximation to $G_{i+\oh}$. 
The use of a larger computational stencil for this purpose can also be justified
by the fact that the derivation of subgrid fluxes using the stochastic homogenization strategy \cite{dta12,dat12,zadoacti18} leads to extended-stencil approximations. Although the homogenization-based approach is valid in the
limit of infinite separation of time scales between the coarse-scale components and fluctuations, a good agreement
between the statistical properties of coarse-mesh variables in the fully resolved model and the reduced model was
obtained in the stationary regime for the Burgers equation \cite{dta12,dat12} and for the SWE \cite{zadoacti18}.
Thus, we seek an approximation 
\[
G_{i+\oh} \approx \tilde{\calG} 
(U_{i-1}, U_i, U_{i+1}, U_{i+2},\omega),
\]
where $\tilde{\calG}$ is a nonlinear function represented by a neural network
and $\omega$ is the vector of neural network's internal parameters (weights and
biases).

\subsection{Network Architecture and Training}
\label{sec:nn}
In this paper, we use a feed-forward neural network as an approximating function 
$\tilde{\calG}$.
We perform direct numerical simulations with the 
high-resolution  model \eqref{eq:semi-discrete-fv} to
generate $M$ data points 
\begin{equation}
\label{data}
(U_{i-1}^{(j)}, U_i^{(j)}, U_{i+1}^{(j)}, U_{i+2}^{(j)}, G_{i+\oh}^{(j)}), \quad j=1,\ldots,M 
\end{equation}
and use the mean square loss function 
\begin{equation}
  \label{eq:mse_loss}
  \mathcal{C}(\omega):=\frac{1}{M}\sum_{j=1}^{M}
  \norm{ \tilde{\mathcal{G}}(U_{i-1}^{(j)}, U_i^{(j)}, U_{i+1}^{(j)}, U_{i+2}^{(j)},\omega)- G_{i+\oh}^{(j)} }_2^2
\end{equation}
to train a neural network. It is important to note that, for any dataset $(U_{i-1}, U_i, U_{i+1}, U_{i+2})$, a fitted approximation to the subgrid flux $G_{i+\oh}$  is not necessarily unique. The ``true'' value \eqref{eq:subgrid_flux_function} of $G_{i+\oh}$ depends on the subgrid states $u_{ki}$ and $u_{ki+1}$, which are treated as unknown.

\begin{figure}[hb]
\centerline{
\includegraphics[scale=1]{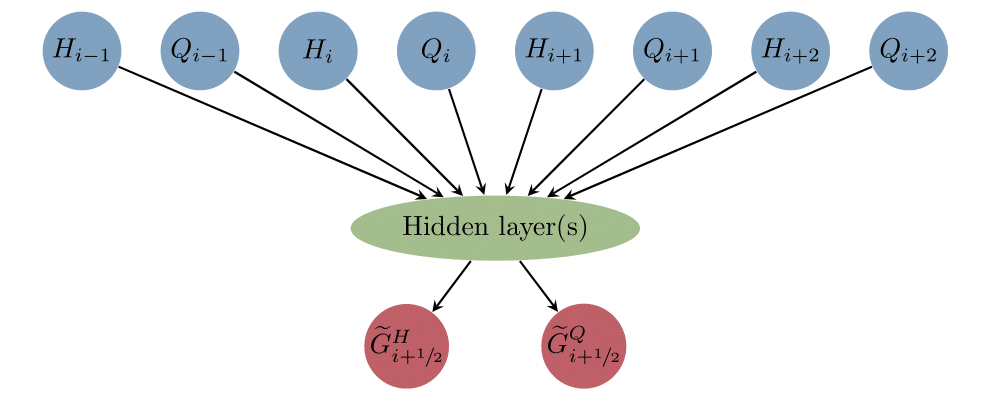}}
\caption{Schematic representation of the neural network used to approximate $G_{i+\oh}=[G_{i+\oh}^H,G_{i+\oh}^Q]$.}
\label{fig0}
\end{figure}

Figure \ref{fig0} represents schematically the neural network used in our computations.
{\mybf 
The neural network has 3 hidden layers with $N_{hid}$ neurons each. 
In this paper we consider 
$N_{hid}=128,64,32,16$ and discuss performance of these 4 neural networks for different parameter regimes.}
We use the leaky ReLU \cite{maas2013rectifier} 
activation function $f(z) = \max\{\alpha z, z\}$ with $\alpha=0.01$.
We also employ early stopping and data normalization as part of our regularization strategy. 
We use the default values of the \verb+NeuralNet+ class of the \verb+skorch+ package, unless stated otherwise. In particular, the 
default training algorithm 
\verb+torch.optim.SGD+ of the \verb+PyTorch+ package is used in our experiments.

For the training dataset,
we generate $n$ trajectories of the SWE using random initial conditions 
\begin{equation}
\label{eq:ic}
h_0(x) = H_0 + A_h \sin(2\pi k_h x/L + \phi_h), \quad 
v_0(x) = V_0 + A_v \sin(2\pi k_v x/L + \phi_v).
\end{equation}
The constant average height $H_0 = 2.0$ is used in all initial conditions for training. The remaining parameters are randomly generated following
the distributions
$V_0 \sim Unif(1, 2)$, $A_h, \, A_v \sim Unif(0.2, 0.6)$,
$k_h, \, k_v \sim Unif\{1, 6\}$, $\phi_h, \, \phi_v \sim Unif(0, 2\pi)$,
where $Unif(a, b)$ denotes the continuous uniform distribution 
on $(a,b)$ and $Unif\{n, m\}$ represents the discrete uniform distribution between $n, m \in N$. We simulate each trajectory until $T^{train}$ and sample with the time step $\Delta t$.
Specific values of $n$, $T^{train}$, and $\Delta t$ are listed in Section \ref{sec:num}.

Next, we run coarse-mesh simulations and compute the ``true'' subgrid fluxes \eqref{eq:subgrid_flux}. In each time step, we generate $N_c$ subgrid fluxes $G_{i+\oh}$, $i=1,\ldots,N_c$. This computation generates $N_c \times n \times T^{train}/\Delta t$ data points for training the neural network with the batch size of 128. 
Initially, we train for 500 epochs with a learning rate of $\gamma = 0.1$ or $\gamma=0.05$.
The learning rate (in this range) for initial training does not seem to affect the quality of overall training at the end.
Subsequently, we continue training for further 2000 epochs with a lower learning rate of $\gamma = 0.001$.
As a regularization technique, we implement
early stopping during both stages of training.

\subsection{Flux Limiting}
\label{sec:mcl}
One of the major drawbacks of machine learning is that there is a considerable gap in the theoretical understanding of neural networks. In our context, 
neural networks are treated as ``black box'' generators of subgrid fluxes
$\tilde G_{i+\oh}$. The addition of these fluxes to $\bar{F}_{i+\oh}$
can produce spurious oscillations and/or
non-physical numerical solutions (e.g., negative water heights). To address this
issue, we integrate the \emph{monolithic convex limiting} (MCL) procedure
\cite{kuzmin2020monolithic} into the flux-corrected coarse-mesh
discretization
\begin{equation}\label{mcl-flux}
  \frac{\d }{\d t}U_i =\frac{(\bar{F}_{i-\oh}+\tilde G_{i-\oh}^*)
  -(\bar{F}_{i+\oh}+\tilde G_{i+\oh}^*)}{\Delta X},
\end{equation}
where $\tilde G_{i\pm\oh}^*$ denotes a constrained approximation to the
network subgrid flux $\tilde G_{i\pm\oh}$. To explain the MCL design
philosophy, we write \eqref{mcl-flux} in the equivalent bar state form
\begin{equation}\label{eq:semi-discrete-llfmcl}
  \frac{\d}{\d t} U_{i}=\frac{1}{\Delta X}\bigl[\Lambda_{i-\oh}(\bar{U}_{i-\oh}^{*,+}-U_i)
    +\Lambda_{i+\oh}(\bar{U}_{i+\oh}^{*,-}-U_i) \bigr],
\end{equation}
where $\Lambda_{i+\oh}$ is the maximum wave speed depending
  on $U_i$ and $U_{i+1}$. The low-order LLF part of
\begin{equation}
\bar{U}_{i+\oh}^{*,\pm}=  \bar{U}_{i+\oh}\pm\frac{\tilde G_{i+\oh}^*}{\Lambda_{i+\oh}}
\end{equation}
is given by
\begin{equation}\label{eq:bar_state-coarse}
  \bar{U}_{i+\oh}:=\frac{U_{i+1}+U_i}{2}-\frac{1}{2\Lambda_{i+\oh}}(f(U_{i+1})-f(U_i)).
\end{equation}

An \emph{invariant domain} of a hyperbolic initial value problem is a convex
admissible set containing all states that an exact solution
may attain \cite{guermond2016invariant}. In the SWE context,
a physical invariant domain
$\mathcal A$ consists of all states $u=[h,hv]^T$ such that the water height
$h$ is nonnegative. The bar state $\bar{U}_{i+\oh}$ represents
an averaged exact solution of the Riemann problem with the
initial states $U_i$ and $U_{i+1}$ \cite{guermond2016invariant,hll1983}. Thus
$\bar{U}_{i+\oh}\in\mathcal A$ provided that $U_i,U_{i+1}\in\mathcal A$.
The MCL approach \cite{hajduk2022algebraically} 
ensures that $\bar{U}_{i\pm\oh}^{*,\mp}\in\mathcal A$
whenever $\bar{U}_{i\pm\oh}\in\mathcal A$. Each forward Euler stage
of Heun's method advances the given data in time
as follows:
\begin{equation}\label{heun}
U_i^{\rm MCL} = U_i+\frac{\Delta T}{\Delta X}\bigl[\Lambda_{i-\oh}(\bar{U}_{i-\oh}^{*,+}-U_i)
    +\Lambda_{i+\oh}(\bar{U}_{i+\oh}^{*,-}-U_i) \bigr].
\end{equation}
If the time step $\Delta T$ satisfies the CFL condition
\begin{equation}\label{cfl}
\frac{\Delta T}{\Delta X}[\Lambda_{i+\oh}+\Lambda_{i-\oh}]\le 1,
\end{equation}
then $U_i^{\rm MCL}$ is a convex combination of $U_i$ and
$\bar{U}_{i\pm\oh}^{*,\mp}$. It follows that (cf.~\cite{guermond2016invariant})
$$
U_i,U_{i\pm1}\in\mathcal A\quad\Rightarrow\quad
U_i,\bar{U}_{i\pm\oh}\in\mathcal A\quad\Rightarrow\quad
U_i,\bar{U}_{i\pm\oh}^{*,\mp}\in\mathcal A\quad\Rightarrow\quad
U_i^{\rm MCL}\in\mathcal A.
$$
Thus the MCL scheme is invariant domain preserving
(positivity preserving) for time steps satisfying
\eqref{cfl}.

\smallskip

The above limiting strategy differs from classical flux-corrected
transport (FCT) algorithms \cite {boris1973,zalesak1979} and their
extensions to hyperbolic systems \cite{guermond2018second,kuzmin2010,lohner1987finite}
in that it does not split the fully discrete scheme
\eqref{heun} into a low-order predictor step
and an antidiffusive corrector step. Instead,
 inequality constraints are imposed on the intermediate states
 $\bar{U}_{i\pm\oh}^{*,\mp}$ of the spatial semi-discretization.
 In addition to constraining them to stay in the set
 $\mathcal A$ of physically admissible states, the MCL
 procedure can
 be configured to enforce semi-discrete entropy inequalities
 and/or local maximum principles for scalar quantities of interest
 \cite{kuzmin-hajduk2022}.
 \smallskip
 
 The \emph{sequential}
 MCL algorithm for the SWE system
 \cite{hajduk2022algebraically,hajduk2022bound} 
  imposes local bounds on $h$ and $v$. In
 the present paper,
 we use a one-dimensional finite volume version of
 this algorithm to limit the fluxes
$$
 \tilde G_{i+\oh}=[\tilde G_{i+\oh}^{h},\tilde G_{i+\oh}^{q}]^T.
 $$
 By abuse of notation, individual
 components of coarse-mesh variables
are denoted by lowercase letters
 in the remainder of this
 section. For example,
  $\bar{U}_{i+\oh}=[\bar h_{i+\oh},\bar h_{i+\oh}\bar v_{i+\oh}]^T$,
 where $\bar v_{i+\oh}:={\bar q_{i+\oh}}/{\bar h_{i+\oh}}
$. Let
 \begin{gather}
   h_i^{\min}:=\min(\bar h_{i-\oh},\bar h_{i+\oh}),
   \qquad h_i^{\max}:=\max(\bar h_{i-\oh},\bar h_{i+\oh}),\\
      v_i^{\min}:=\min(\bar v_{i-\oh},\bar v_{i+\oh}),
   \qquad v_i^{\max}:=\max(\bar v_{i-\oh},\bar v_{i+\oh}).
   \end{gather}
Following Hajduk \cite{hajduk2022algebraically}, we formulate 
local maximum principles for the limited bar states as follows:
\begin{subequations}\label{mcl-seq}
 \begin{gather}\label{mcl-seq-h}
   h_i^{\min}\le \bar h_{i+\oh}^{*,-}\le h_i^{\max},\qquad
   h_{i+1}^{\min}\le \bar h_{i+\oh}^{*,+}\le h_{i+1}^{\max},\\
   \label{mcl-seq-q}
   \bar h_{i+\oh}^{*,-}
   v_i^{\min}\le \bar q_{i+\oh}^{*,-}\le
\bar h_{i+\oh}^{*,-}
v_i^{\max},\qquad
\bar h_{i+\oh}^{*,+}
   v_{i+1}^{\min}\le \bar q_{i+\oh}^{*,+}\le \bar h_{i+\oh}^{*,+}v_{i+1}^{\max}.
 \end{gather}
\end{subequations}
Since $\bar h_{i+\oh}^{*,\pm}=\bar h_{i+\oh}\pm
\frac{\tilde G_{i+\oh}^{h,*}}{\Lambda_{i+\oh}}$, the water height
constraints \eqref{mcl-seq-h} are equivalent to
\begin{gather}
\Lambda_{i+\oh}(\bar h_{i+\oh}-h_i^{\max})\le 
\tilde G_{i+\oh}^{h,*}\le\Lambda_{i+\oh}(\bar h_{i+\oh}-h_i^{\min}),\\
\Lambda_{i+\oh}(h_{i+1}^{\min}-\bar h_{i+\oh})\le
\tilde G_{i+\oh}^{h,*}\le
\Lambda_{i+\oh}(h_{i+1}^{\max}-\bar h_{i+\oh}).
\end{gather}
It follows that the best bound-preserving approximation to
$\tilde G_{i+\oh}^{h}$ is defined by
\begin{equation}\label{mcl-limiter-h}
  \tilde G_{i+\oh}^{h,*}=\begin{cases}
  \min(  \tilde G_{i+\oh}^{h},\Lambda_{i+\oh}\min(
  \bar h_{i+\oh}-h_i^{\min},h_{i+1}^{\max}-\bar h_{i+\oh}))
  & \mbox{if}\  \tilde G_{i+\oh}^{h}\ge 0,\\
  \max(  \tilde G_{i+\oh}^{h},\Lambda_{i+\oh}\max(
  \bar h_{i+\oh}-h_i^{\max},h_{i+1}^{\min}-\bar h_{i+\oh}))
  & \mbox{otherwise}.
  \end{cases}
\end{equation}  

To ensure the feasibility of the discharge constraints
\eqref{mcl-seq-q}, the bar state
\[
\bar q_{i+\oh}^{*,\pm}=\bar q_{i+\oh}\pm\frac{
  \tilde G_{i+\oh}^{q,*}}{\Lambda_{i+\oh}}=
\bar h_{i+\oh}^{*,\pm}\bar v_{i+\oh}\pm\frac{\Delta
  \tilde G_{i+\oh}^{q,*}}{\Lambda_{i+\oh}}
\]
is defined using a limited counterpart
$
  \Delta\tilde G_{i+\oh}^{q,*}= \tilde G_{i+\oh}^{q,*}-\tilde G_{i+\oh}^{h,*}
  \bar v_{i+\oh}
  $
  of the flux difference
  \[
  \Delta\tilde G_{i+\oh}^{q}= \tilde G_{i+\oh}^{q}-\tilde G_{i+\oh}^{h,*} \bar v_{i+\oh}.
  \]
  The sequential MCL algorithm \cite{hajduk2022algebraically} calculates the
  limited differences $ \Delta\tilde G_{i+\oh}^{q,*}$ and sets
  \begin{equation}\label{mcl-split}
  \tilde G_{i+\oh}^{q,*}:=\tilde G_{i+\oh}^{h,*}
  \bar v_{i+\oh}+\Delta\tilde G_{i+\oh}^{q,*}.
  \end{equation}
  Similarly to the derivation of the water height limiter
  \eqref{mcl-limiter-h}, we use the equivalent form
  \begin{gather}
\Lambda_{i+\oh}\bar h_{i+\oh}^{*,-}(\bar v_{i+\oh}-v_i^{\max})\le 
\Delta \tilde G_{i+\oh}^{q,*}\le\Lambda_{i+\oh}\bar h_{i+\oh}^{*,-}
(\bar v_{i+\oh}-v_i^{\min}),\\
\Lambda_{i+\oh}\bar h_{i+\oh}^{*,+}(v_{i+1}^{\min}-\bar v_{i+\oh})\le
\Delta\tilde G_{i+\oh}^{q,*}\le
\Lambda_{i+\oh}\bar h_{i+\oh}^{*,+}(v_{i+1}^{\max}-\bar v_{i+\oh})
  \end{gather}
  of conditions \eqref{mcl-seq-q}
  to establish their validity for \eqref{mcl-split}
  with $\Delta\tilde G_{i+\oh}^{q,*}$ defined by 
\begin{equation}\label{mcl-limiter-q}
  \Delta \tilde G_{i+\oh}^{q,*}=\begin{cases}
  \min(  \Delta\tilde G_{i+\oh}^{q},\Lambda_{i+\oh}\min(
  \bar h_{i+\oh}^{*,-}(\bar v_{i+\oh}-v_i^{\min}),
  \bar h_{i+\oh}^{*,+}(v_{i+1}^{\max}-\bar v_{i+\oh})))
  & \mbox{if}\  \Delta \tilde G_{i+\oh}^{q}\ge 0,\\
  \max(\Delta  \tilde G_{i+\oh}^{q},\Lambda_{i+\oh}\max(
  \bar h_{i+\oh}^{*,-}(\bar v_{i+\oh}-v_i^{\max}),
  \bar h_{i+\oh}^{*,+}(v_{i+1}^{\min}-\bar v_{i+\oh})))
  & \mbox{otherwise}.
  \end{cases}
\end{equation}  
  
The enforcement of \eqref{mcl-seq} guarantees
positivity preservation and numerical admissibility
of approximate solutions to the one-dimensional SWE system.
 For a more detailed description of the sequential MCL
 algorithm, we refer
 the interested reader to Schwarzmann \cite{alexey_thesis}.
 Other examples of modern flux correction schemes for nonlinear hyperbolic
systems and, in particular, for SWEs can be found
in \cite{azerad2017,kuzmin2010,lohmann2016synchronized,guermond2018second,pazner2021sparse}.

\section{Numerical Results}
\label{sec:num}
Considerable effort has been made to develop NN-based numerical approximations that are consistent with the physics of underlying PDE models, but the analysis of the resulting numerical algorithms can be difficult. 
Thus, 
although deep learning has become extremely popular in the context of computational methods for PDEs, numerical schemes based on neural networks can produce non-physical solutions. 
In the approach that we proposed above, subgrid fluxes generated by a neural network are ``controlled'' by a mathematically rigorous flux limiting formalism.
The main goal here is to demonstrate that flux limiting can improve the quality of a deficient subgrid model by eliminating non-physical oscillations.

In our fine-mesh simulations for the SWE system \eqref{eq:swe_conservative_form} with  $g = 9.812$, we  discretize the computational domain $\Omega=(0,100)$ using $N_f=2000$ cells. As before, we assume that the fine-mesh discretization 
resolves all physical processes of interest properly. In particular, the numerical diffusion is negligible and further refinement does not produce any visible changes in the  numerical solutions.  We perform
fine-mesh simulations using the time step $\Delta t = 0.1\Delta x$. The
IDP property of the LLF space discretization and the
SSP property of Heun's time-stepping method guarantee positivity
preservation for this choice of $\Delta t$.
Using the coarsening parameter  $k=20$, we construct coarse-mesh discretizations
with $N_c=100$ and $\Delta X = 1$.
{\mybf
The time step $\Delta T=0.005$
of the reduced model is an integer multiple of $\Delta t$.
We also investigate the performance of reduced coarse models integrated numerically with larger time-steps $\Delta T=0.0$ and $\Delta T=0.1$.
We consider the NN subgrid model with $N_{hid}=128$ first, and then compare the performance 
of this NN with NNs with $N_{hid}=64$, $32$, $16$ in a later paragraph.}
Initial conditions for the testing dataset are generated using equation \eqref{eq:ic} with  parameter values that are not present in the training dataset. Selected parameters of initial conditions used for testing are presented in Table \ref{tab:ic}.
The software used in this paper is available on GitHub
\cite{Schwarzmann_clf1_A_collection}.

We tested many initial conditions in this study. The overall performance of our subgrid model is discussed at the end of this section. We refer to simulations on the fine mesh as \emph{full model} experiments or DNS
(Direct Numerical Simulation). The results of reduced model experiments are referred to as \emph{NN-reduced} if no MCL flux correction is performed and as \emph{MCL-NN-reduced} if the MCL fix is activated.
%
\begin{table}[h]
  \centering
  \begin{tabular}{c|llll|llll}
    \hline
    IC & $H_0$ & $A_h$ & $k_h$ & $\varphi_h$
    & $V_0$ & $A_v$ & $k_v$ & $\varphi_v$ \\
    \hline
    1 & 2.0 & 0.2 & 3 & 0.0 & 1.0 & 0.0 & 0 & 0.0 \\
    2 & 2.0 & 0.45 & 4 & 2.78 & 1.1 & 0.5 & 3 & 4.5 \\
    3 & 2.0 & 0.45 & 4 & 2.78 & 1.1 & 0.5 & 4 & 4.5 \\
    4 & 2.0 & 0.2 & 3 & 0.0 & 1.1 & 0.1 & 2 & 0.0 \\
    \hline
  \end{tabular}  
  \caption{Parameters of initial conditions \eqref{eq:ic} used for testing.}
  \label{tab:ic}
\end{table}

A visual comparison of numerical solutions to the SWE obtained with $N_f = 2000$ and $N_c = N_f / k$ confirms that the subgrid fluxes $G_{i+\oh}$ are antidiffusive (not shown here; 
see \cite{alexey_thesis}). Therefore, subgrid fluxes should ``remove'' additional diffusion resulting from the coarse-mesh discretization. However, this process is nonlinear and quite delicate. Since there is no control over the numerical performance of the neural network used to model subgrid fluxes, the NN-reduced model might produce non-physical solutions.

Several networks with the same configuration but with different amounts of training data are tested in this study.
The amount of training data is determined by two parameters: (i) $n$, how many different trajectories are generated for training and 
(ii) $T^{train}$, the length of each trajectory. Each trajectory is sampled with the time step $\Delta t=0.1\Delta x$.
Given a lot of training data, the network typically reproduces fluxes quite well and the reduced model performs much better. 
However, in practical situations, the amount of training data can be limited since it is impossible to
know a priori how many trajectories need to be generated for training.

\begin{figure}[ht]
\centerline{\includegraphics[scale=0.5]{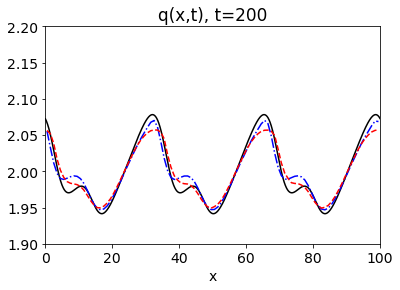}
\includegraphics[scale=0.5]{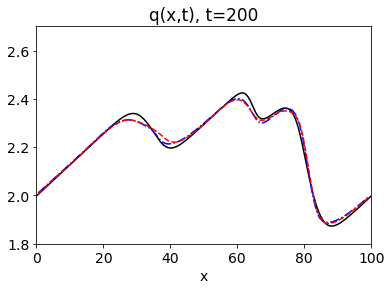}}
\centerline{\includegraphics[scale=0.5]{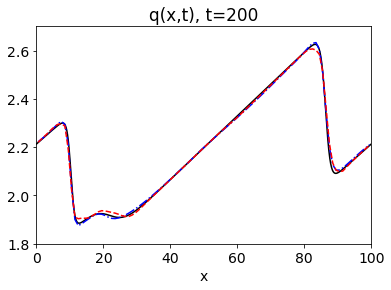}
\includegraphics[scale=0.5]{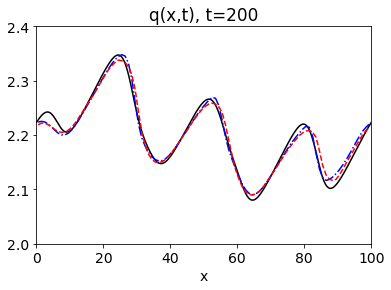}}
\caption{Discharge $q(x,t)=hv$ at time $t=200$ in simulations of the full model (black solid line) with $\Delta x=0.05$ and reduced models with $k=20$ ($\Delta x=1$) and NN with $N_{hid}=128$ trained with $n=200$ and $T=80$;  Blue Dash-Dot Line - NN-reduced model,  
 Red Dashed Line - MCL-NN-reduced model.
Initial conditions with parameters in Table \ref{tab:ic} (1 - top-left, 2 - top-right, 3 - bottom-left, 4 - bottom-right).}
\label{fig1}
\end{figure}

\begin{figure}[ht]
\centerline{\includegraphics[scale=0.5]{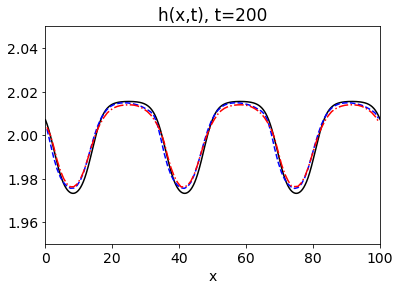}
\includegraphics[scale=0.5]{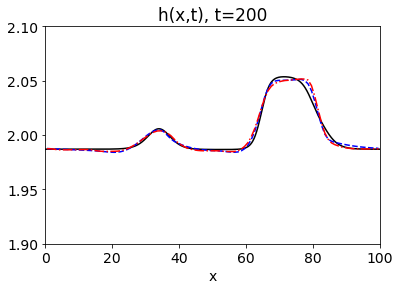}}
\centerline{\includegraphics[scale=0.5]{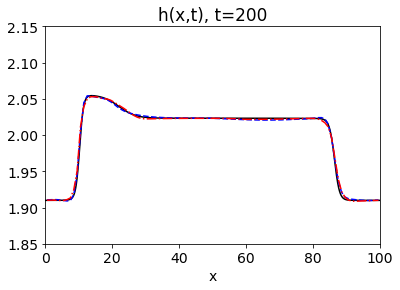}
\includegraphics[scale=0.5]{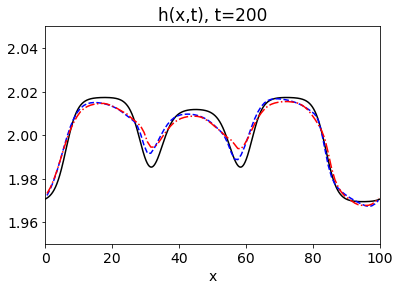}}
\caption{Water height $h(x,t)$ at time $t=200$ in simulations of the full model (black solid line) with $\Delta x=0.05$ and reduced models with $k=20$ ($\Delta x=1$) and NN with $N_{hid}=128$ trained with $n=200$ and $T=80$;  Blue Dash-Dot Line - NN-reduced model,  
 Red Dashed Line - MCL-NN-reduced model.
Initial conditions with parameters in Table \ref{tab:ic} (same location as in Figure \ref{fig1}).}
\label{fig2}
\end{figure}

First, we present results for a relatively large amount of training data with 
$n=200$ and $T^{train}=80$.
Figures \ref{fig1} and \ref{fig2} depict the discharge $q = hv$ and water height $h$ at time $t=200$
for several choices of initial conditions in simulations with the full model and both reduced models (without and with limiting).
%
%
%

\begin{figure}[ht]
\centerline{\includegraphics[scale=0.5]{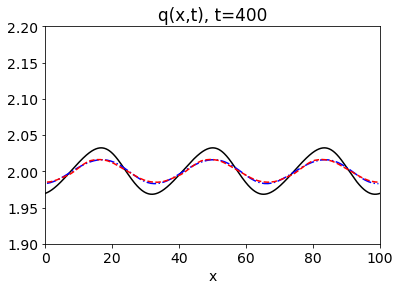}
\includegraphics[scale=0.5]{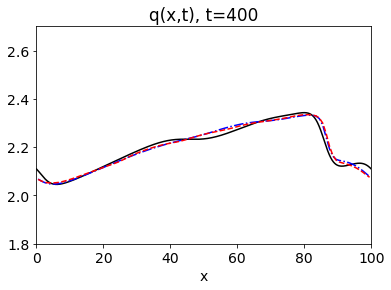}}
\centerline{\includegraphics[scale=0.5]{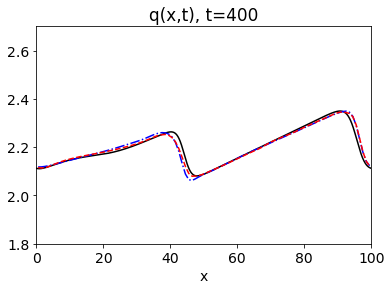}
\includegraphics[scale=0.5]{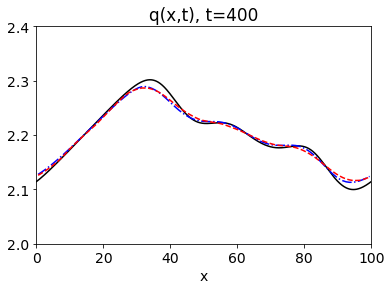}}
\caption{Discharge $q(x,t)=hv$ at time $t=400$ in simulations of the full model (black solid line) with $\Delta x=0.05$ and reduced models with $k=20$ ($\Delta x=1$) and NN with $N_{hid}=128$ trained with $n=200$ and $T=80$;  Blue Dash-Dot Line - NN-reduced model,  
 Red Dashed Line - MCL-NN-reduced model.
Initial conditions with parameters in Table \ref{tab:ic} (same location as in Figure \ref{fig1}). Notice the difference in the vertical scale between the upper left subplot and the other three subplots.}
\label{fig3}
\end{figure}

In this example, the application of MCL to the subgrid fluxes does not significantly affect the quality of the results. It is worth pointing out that we test the performance of reduced models on a much longer interval compared to the length of training trajectories. Specifically, the length of the testing trajectory is $T^{test}=200$, while the length of each training trajectory is $T^{train} = 80$. 
The performance of
reduced models deteriorates for larger times and solutions with relatively small waves, but our ML reduced models can predict the dynamics of the SWE with larger waves for a much longer time. Numerical solutions for $T^{test} = 400$ are compared in Figure \ref{fig3}.

\begin{figure}[ht]
\centerline{\includegraphics[scale=0.5]{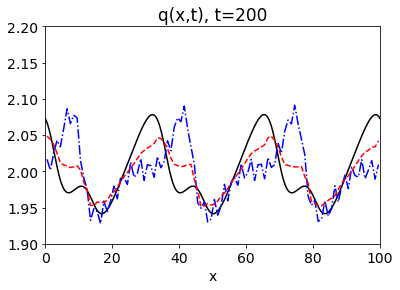}
\includegraphics[scale=0.5]{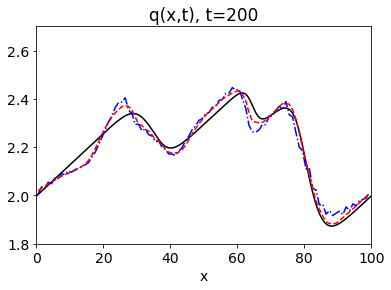}}
\centerline{\includegraphics[scale=0.5]{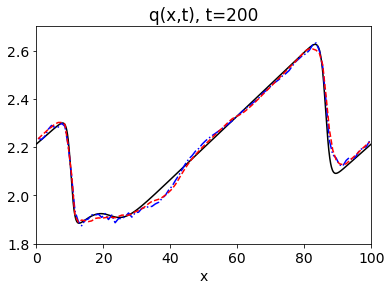}
\includegraphics[scale=0.5]{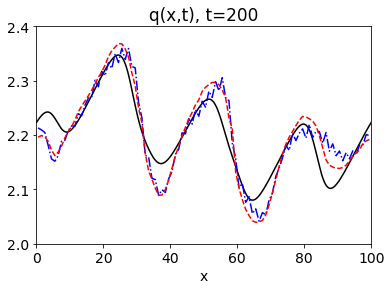}}
\caption{Discharge $q(x,t)=hv$ at time $t=200$ in simulations of the full model (black solid line) with $\Delta x=0.05$ and reduced models with $k=20$ ($\Delta x=1$) and NN with $N_{hid}=128$ trained with $n=100$ and $T^{train}=40$;  Blue Dash-Dot Line - NN-reduced model,  
 Red Dashed Line - MCL-NN-reduced model.
Initial conditions with parameters in Table \ref{tab:ic} (same location as in Figure \ref{fig1}).}
\label{fig4}
\end{figure}

The performance of ML reduced models changes drastically when the network is trained on a smaller dataset. For instance, we also generated the training dataset with 
$n=100$ and $T^{train}=40$.
Thus, the training dataset is four times smaller compared to the training dataset for the reduced model used above. 
A comparison of the discharge at time $T=200$
for the DNS and simulations with the two reduced models (without and with the MCL fix) is presented in Figure \ref{fig4}. The reduced model without the limiter produces spurious oscillations in all four cases. These oscillations are particularly evident in the top-left subplot. The MCL-NN reduced model corrects these small-scale oscillations and produces physical solutions.

One can use the MCL limiter as an indicator of the training quality for the NN reduced model. The MCL limiter is applied after training and the MCL-NN reduced model has approximately the same computational cost as the NN reduced model without the limiter. Therefore, if the two models produce very similar results, then this implies that the neural network approximates the fluxes with very good accuracy and is unlikely to produce non-physical solution values. However, if the two reduced models result in considerably different numerical solutions, then this can be an indicator that the neural network is not trained sufficiently well.

{\mybf
\textbf{Comparison with Traditional Subgrid Models.}
We also compared our subgrid closure with more traditional approaches for modeling the subgrid terms. In particular, 
the application of a flux limiter to a coarse-mesh discretization in conjunction with a high-order flux approximation can be interpreted as an implicit subgrid scale model for ``monotonically integrated'' Large Eddy Simulation (MILES); see, e.g., \cite{miles,miles1,miles2}. Thus, we compare our subgrid NN closure with MCL-constrained results for
target fluxes corresponding to the Lax-Wendroff closure.

Short simulations reveal that the MCL-Lax-Wendroff (MCL-LW) scheme is much more diffusive compared to our MCL-NN closure model. Numerical results comparing the two subgrid models are depicted in Figure~\ref{fig5}. This figure demonstrates that the MCL-LW model with $N_x=100$ ($\Delta x=1$) is very diffusive and does not represent the accurate numerical solution at all.
The MCL-LW model with $N_x=500$ ($\Delta x=0.2$) is also diffusive and leads to incorrect (higher) minima and considerably less sharp fronts in numerical solutions.
Our MCL-NN subgrid model with $N_x=100$ ($\Delta x=1$) correctly represents the numerical solution in resolved simulations. We also verified these results with different initial conditions and longer simulations times. They are not presented here only for the brevity of the presentation.
\begin{figure}[ht]
\centerline{\includegraphics[scale=0.75]{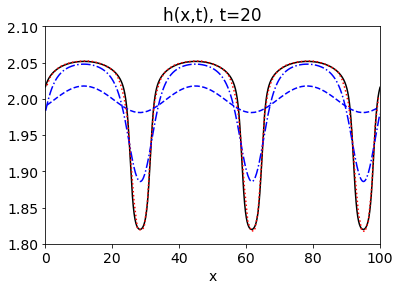}}
\caption{Water height $h(x,t)$ at time $t=20$ in simulations of the full model (black solid line) with $N_x=2000$ ($\Delta x=0.05$), MCL-LW models with $N_{hid}=128$, $N_x=100$ ($\Delta x=1$) and $N_x=500$ ($\Delta x=0.2$) (blue dashed and blue dash-dot lines, respectively), and MCL-NN reduced models with $N_x=100$ ($\Delta x=1$) trained with $n=200$ and $T^{train}=80$ (red dotted line nearly overlaps with the black line). Initial conditions \#1 in Table~\ref{tab:ic}.}
\label{fig5}
\end{figure}

\textbf{Simulations with MCL-NN Subgrid Model using Different Time-Steps.}
The CFL requirement for the fully resolved model appears to be approximately $\Delta t = 0.1 \Delta x$ where $\Delta x=L/N_f = 0.05$.
The upper bound on the wave propagation is given by \eqref{eq:wave_speed}, 
and $\sqrt{gH_0} \approx 4.4$. Initially, $V_0 \approx 1$ and the upper bound on the speed of wave propagation can be roughly estimated as $|V_0| + \sqrt{gH_0} \approx 5.4$. 
Thus, the LLF model with $N_f=2000$ and $\Delta t = 0.2 \Delta x$ violates the CFL condition and produces non-physical oscillations (not shown here).

The main advantage of the coarse model is that it allows using a much larger time-step, since the mesh size is significantly larger than in the fully resolved simulations. Figures \ref{fig6} and \ref{fig7} depict a comparison of simulations for MCL-NN models integrated with several different time-steps $\Delta t = 0.005$, $0.05$, $0.1$. Recall that the time-step of the fully resolved LLF model is $\Delta t=0.005$. There are small discrepancies between MCL-NN reduced models integrated with different time-steps, but numerical results with three different time-steps are in a very good agreement with fully resolved simulations. Thus, coarser mesh simulations with neural networks allow for additional acceleration due to a much larger time-step compared with the fully resolved model. 
\begin{figure}[ht]
\centerline{\includegraphics[scale=0.4]{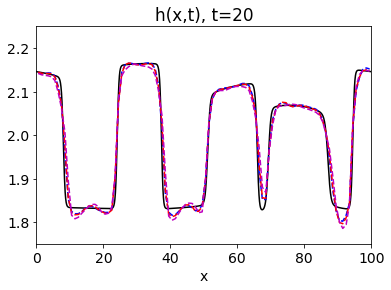}
\includegraphics[scale=0.4]{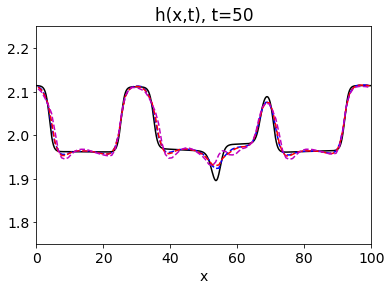}
\includegraphics[scale=0.4]{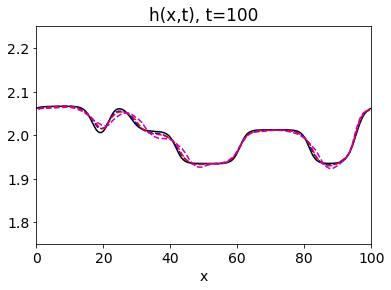}}
\centerline{\includegraphics[scale=0.4]{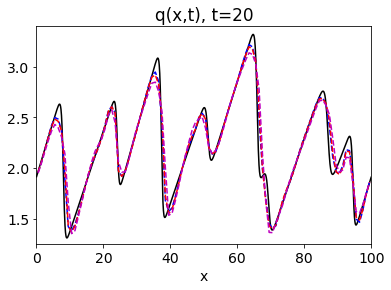}
\includegraphics[scale=0.4]{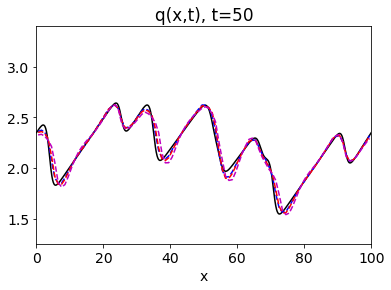}
\includegraphics[scale=0.4]{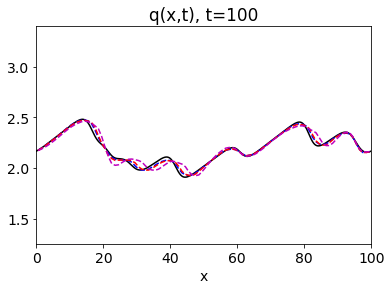}}
\caption{Water height $h(x,t)$ and discharge $q(x,t)$ at time $t=20$, $50$, $100$ in simulations of the full model (black solid line) and MCL-NN reduced models with $N_{hid}=128$, $N_x=100$ ($\Delta x=1$) trained with $n=200$ and $T^{train}=80$ 
with time-step $\Delta t = 0.005$, $0.05$, $0.1$
(dashed blue, red, and magenta lines, respectively). 
Initial conditions \#2 in Table \ref{tab:ic}.}
\label{fig6}
\end{figure}
\begin{figure}[ht]
\centerline{\includegraphics[scale=0.4]{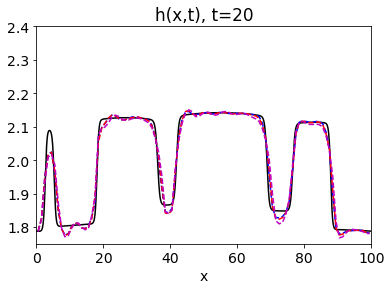}
\includegraphics[scale=0.4]{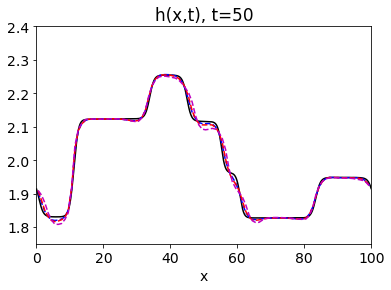}
\includegraphics[scale=0.4]{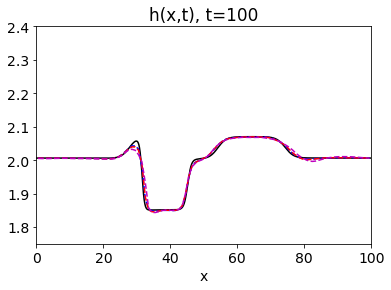}}
\centerline{\includegraphics[scale=0.4]{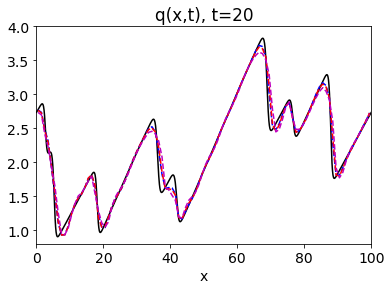}
\includegraphics[scale=0.4]{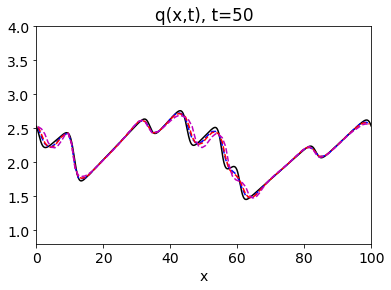}
\includegraphics[scale=0.4]{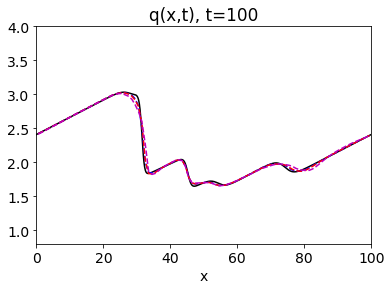}}
\caption{Water height $h(x,t)$ and discharge $q(x,t)$ at time $t=20$, $50$, $100$ in simulations of the full model (black solid line) and MCL-NN reduced models with $N_{hid}=128$, $N_x=100$ ($\Delta x=1$) trained with $n=200$ and $T^{train}=80$ with time-step $\Delta t = 0.005$, $0.05$, $0.1$
(dashed blue, red, and magenta lines, respectively). 
Initial conditions \#3 in Table \ref{tab:ic}.}
\label{fig7}
\end{figure}

\textbf{Simulations with Networks of Different Sizes with the Same Training Data.}
Since accelerated simulations are the main motivation for introducing subgrid models, we investigate the performance of neural networks with decreasing number of neurons in hidden layers. The overall architecture of neural networks remains the same (i.e. 3 hidden layers) as discussed in section \ref{sec:nn}, but we consider NNs with $N_{hid}=128$, $64$, $32$, $16$
neurons in each hidden layer. Recall that the neural network discussed previously has 3 hidden layers with $N_{hid}=128$ neurons in each layer. 
We train all 4 neural networks on the same data set with 
$n=200$ and $T^{train}=80$. In this paragraph, we describe the performance of 4 neural networks in the regime that is close to the training data. In particular, we consider initial conditions such that $H_0 \approx 2$ and $V_0 \approx 1$.
We also use the MCL limiter in all simulations (we refer to these closure models as MCL-NN).
Figure \ref{fig8}
depicts a comparison of the water heights $h(x,t)$ for the fully resolved simulations and simulations with 4 neural networks using the time-step $\Delta t=0.05$. Velocity and discharge follow a similar trend. Overall, the performance of  3 neural networks 
with $N_{hid}=128, 64, 32$
is very good. In fact, the performance of MCL-NN with $N_{hid}=32$ is sometimes better
than that of MCL-NN with $N_{hid}=128$. This might occur because the neural network with $N_{hid}=128$ has too many parameters and is overfitted. The training loss is increasing with $N_{hid}$. In particular, at the end of training, the training loss for $N_{hid}=128, 64, 32, 16$
is $loss = 1.5, 1.8, 2.7, 6.4 \, (\times 10^{-4})$, respectively.
It it noticeable that the performance of MCL-NN with $N_{hid}=16$  deteriorates for longer times $t>150$.
The performance of MCL-NN with $N_{hid}=16$ is comparable with other networks for times $t<100$. However, using NN with $N_{hid}=16$ without the MCL
sometimes leads to non-physical oscillatory solutions. Thus, the MCL limiter becomes essential for correcting these oscillations, and simulations of the MCL-NN subgrid model agree with the resolved simulations reasonably well.
The oscillatory behavior of NN with $N_{hid}=16$ and corrected MCL-NN with $N_{hid}=16$ is depicted in Figure \ref{fig9}.

We can also observe that all four MCL-NN subgrid models produce small low-frequency oscillations near fronts with high gradients. This is particularly visible for shorter times (e.g. Figure \ref{fig9}). However, due to the MCL limiter, these oscillations do not grow in time, and the coarse mesh solutions agree with the results of fully-resolved simulations very well. In addition, we also observe that the subgrid MCL-NN model with $N_{hid}=32$ performs slightly better than other models. 
Running times are presented in Appendix \ref{app:rt}. PyTorch automatically parallelizes simulations with MCL-NN128 and MCL-NN64. Running times reported in the appendix take the CPU utilization into account and are equivalent to a single-CPU simulation. Parallelized NN simulations are considerably shorter (in real time) than a cingle-CPU fully resolved simulation. However, parallelization might introduce a considerable additional overhead; thus running times presented here serve only as a rough guideline.
\begin{figure}[ht]
\centerline{\includegraphics[scale=0.5]{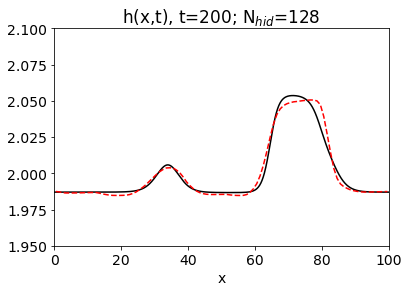}
\includegraphics[scale=0.5]{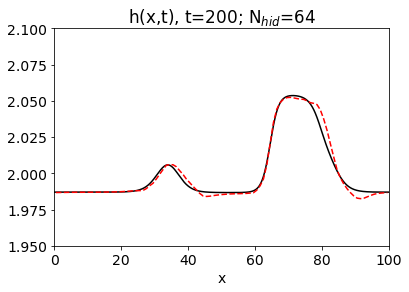}}
\centerline{\includegraphics[scale=0.5]{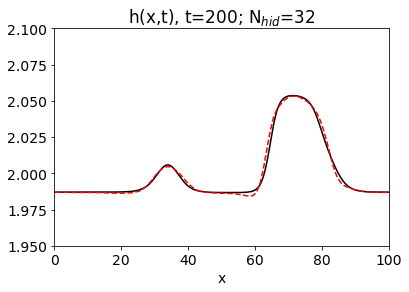}
\includegraphics[scale=0.5]{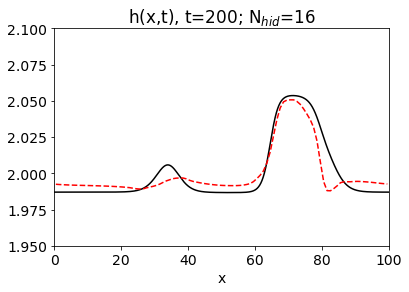}}
\caption{Water height $h(x,t)$ at time $t=200$ in simulations of the full model (black solid line) and MCL-NN reduced models with $N_x=100$ ($\Delta x=1$) trained with $n=200$ and $T^{train}=80$ with time-step $\Delta t = 0.05$
(red dashed line). 
Top left, top right, bottom left, bottom right - 
$N_{hid}=128, 64, 32, 16$, respectively.
Initial conditions \#2 in Table \ref{tab:ic}.}
\label{fig8}
\end{figure}
\begin{figure}[ht]
\centerline{\includegraphics[scale=0.38]{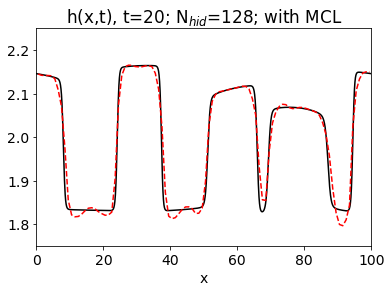}
\includegraphics[scale=0.38]{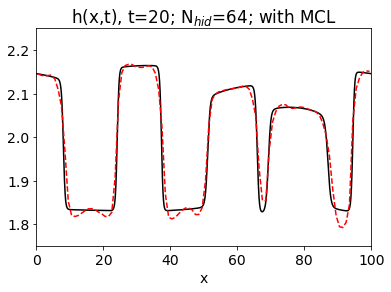}
\includegraphics[scale=0.38]{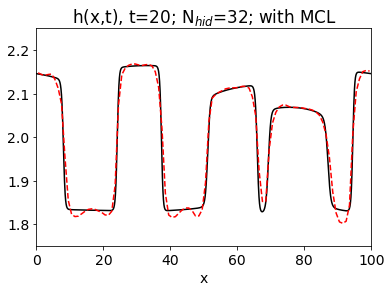}}
\centerline{\includegraphics[scale=0.38]{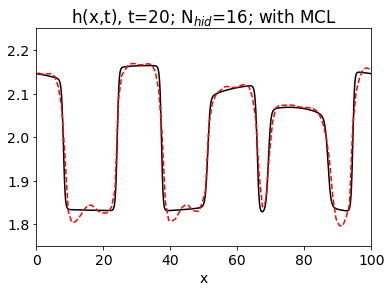}
\includegraphics[scale=0.38]{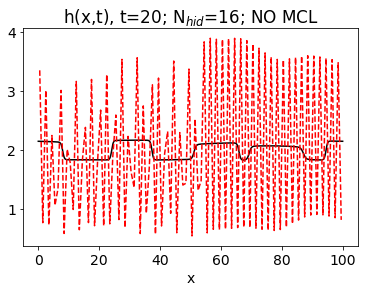}}
\caption{Water height $h(x,t)$ at time $t=20$ in simulations of the full model (black solid line) and NN reduced models with $N_x=100$ ($\Delta x=1$) trained with $n=200$ and $T^{train}=80$ with time-step $\Delta t = 0.05$ (red dashed line). 
Top row -  $N_{hid}=128, 64, 32$ with the MCL.
Bottom left - $N_{hid}=16$ with the MCL, bottom right - $N_{hid}=16$ without the MCL.
Initial conditions \#2 in Table \ref{tab:ic}.}
\label{fig9}
\end{figure}

\textbf{Simulations with Networks of Different Sizes outside of the Training Regime.}
Figures \ref{fig10} and~\ref{fig11} depict simulation results for reduced MCL-NN models
with $N_{hid}=128, 64, 32, 16$ outside of the training regime. In particular, all four subgrid models are trained in the regime $H_0=2$  and then tested with the averaged water height $H_0=1.5$ and $H_0=2.5$. This is a $\pm$25\% fluctuation of $H_0$ compared to the training regime.
Here we present results for only one initial condition and two values of $H_0$, but we also tested on other initial conditions and values of parameters with comparable results.
All four subgrid models perform well outside of the training regime, except, possibly, the MCL-NN model with $N_{hid}=16$ for $H_0=1.5$.
Moreover, the MCL-NN with $N_{hid}=32$ performs slightly better than 
the MCL-NN subgrid model with $N_{hid}=128$ and the MCL-NN subgrid model with  $N_{hid}=64$ produces the best agreement with the fully-resolved simulations overall.
Performance of all four MCL-NN models deteriorates for values of $H_0$ outside of the $\pm$25\% range. The performance of MCL-NN subgrid models declines faster for smaller values of $H_0$. There is still a reasonably good agreement between coarse models and fully resolved simulations for larger values of $H_0 \in [2.5,\ldots,3]$. We obtained similar results for simulations using coarse models with a different averaged velocity in the range $V_0 \in [0.6,\ldots,1.4]$. Performance of subgrid MCL-NN models is less sensitive to fluctuations in $V_0$, probably because $V_0$ is not a conserved quantity. Moreover, the training dataset is generated with $V_0 \in [1,2]$. 
Thus, $V_0$ has much larger (relative) fluctuations in the training dataset compared to $H_0$.
Overall, MCL-NN subgrid models with $N_{hid}=128$, $64$, $32$ perform very well outside of the training regime for a wide range of parameters $H_0$ and $V_0$. We also observe that in tests with varying $H_0$ and $V_0$, the
MCL-NN subgrid models with $N_{hid}=64, 32$ perform comparably to the MCL-NN subgrid model with  $N_{hid}=128$. This is another indication that the computational complexity of MCL-NN subgrid models can be reduced significantly compared to the computational cost of NN with  $N_{hid}=128$.
\begin{figure}[ht]
\centerline{\includegraphics[scale=0.5]{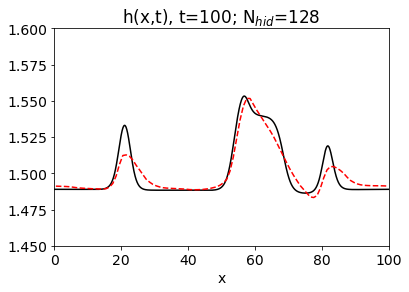}
\includegraphics[scale=0.5]{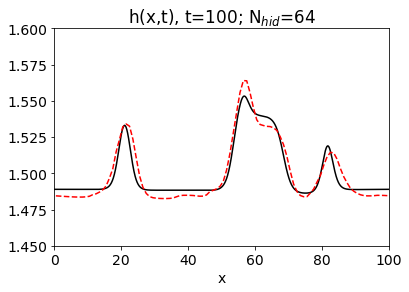}}
\centerline{\includegraphics[scale=0.5]{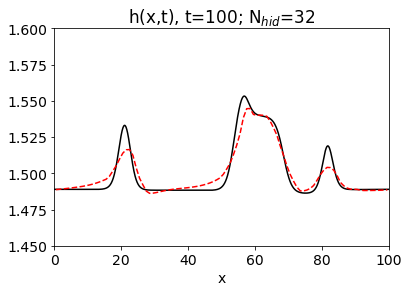}
\includegraphics[scale=0.5]{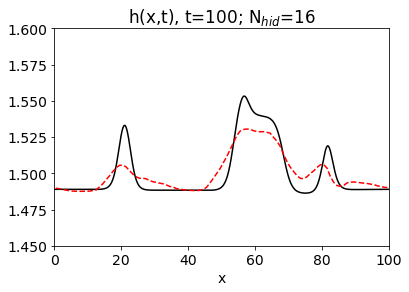}}
\caption{Water height $h(x,t)$ at time $t=100$ in simulations of the full model (black solid line) with $H_0=1.5$ and MCL-NN reduced models with $N_x=100$ ($\Delta x=1$) trained on $H_0=2$ with $n=200$ and $T^{train}=80$ with time-step $\Delta t = 0.05$
(red dashed line). 
Top left, top right, bottom left, bottom right - 
$N_{hid}=128, 64, 32, 16$, respectively.
Initial conditions with $H_0=1.5$ and other parameters as in \#2 in Table \ref{tab:ic}.
Compare with the top right plot in Figure \ref{fig6} where $H_0=2$.
}
\label{fig10}
\end{figure}
\begin{figure}[ht]
\centerline{\includegraphics[scale=0.5]{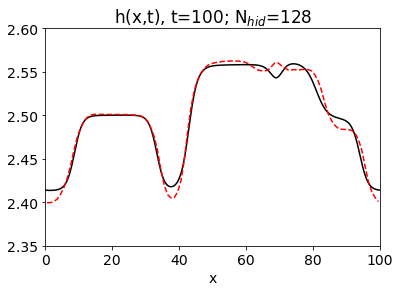}
\includegraphics[scale=0.5]{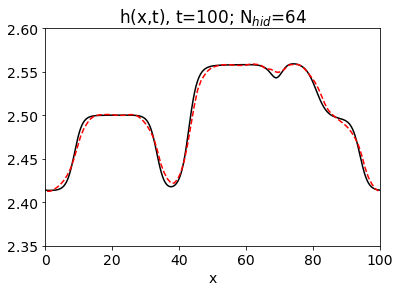}}
\centerline{\includegraphics[scale=0.5]{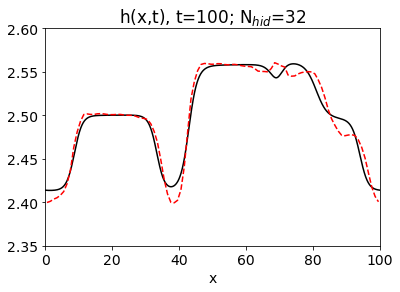}
\includegraphics[scale=0.5]{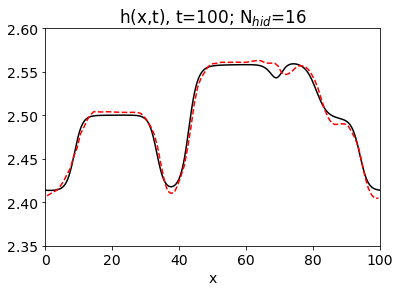}}
\caption{Water height $h(x,t)$ at time $t=100$ in simulations of the full model (black solid line) with $H_0=2.5$ and MCL-NN reduced models with $N_x=100$ ($\Delta x=1$) trained on $H_0=2$ with $n=200$ and $T^{train}=80$ with time-step $\Delta t = 0.05$
(red dashed line). 
Top left, top right, bottom left, bottom right - 
$N_{hid}=128, 64, 32, 16$, respectively.
Initial conditions with $H_0=2.5$ and other parameters as in \#2 in Table \ref{tab:ic}.
Compare with the top right plot in Figure \ref{fig6} where $H_0=2$.
}
\label{fig11}
\end{figure}

\textbf{Simulations with Non-Smooth Initial Conditions.}
We also consider the initial condition
\begin{equation}
\label{damic}
h(x,0) = 
\begin{cases}
    H_0 + a & \frac{L}{3} < x < \frac{2L}{3}, \\
    H_0 - a/2 & \text{otherwise}
\end{cases}, \qquad v(x,0) = 0,
\end{equation}
in which
$H_0$ represents the averaged water height, and $a$ controls the magnitude of the ``step.'' In particular, we tested the performance of subgrid models discussed in previous sections 
with $H_0=1.6$, $2$, $2.4$ and $a=0.5$, $0.75$ for the initial condition given by \eqref{damic}. Similarly to other simulations presented earlier, we tested subgrid NN models with 
$N_{hid}=128$, $64$, $32$, $16$ trained on $n=200$ and $T^{train} = 80$.
We would like to point out that all subgrid NN models are trained on smooth data with $H_0=2$.
The MCL limiter is crucial in simulations with non-smooth initial conditions. Without the limiter, all subgrid NN models quickly develop high-frequency spurious oscillations. The MCL limiter nearly eliminates these oscillations. Numerical results for the water height in simulations with initial \eqref{damic} with $H_0=2.4$ and $a=0.5$ for subgrid models with 
$N_{hid}=128$ and $N_{hid}=32$
are presented in Figure \ref{fig12}. We obtained similar results 
for subgrid models with $N_{hid}=64$ and $N_{hid}=16$.
We would like to note that numerical all solutions of coarse models have small oscillations, but these oscillations do not grow in time due to the MCL. Overall, these results are consistent with our intuition from numerical analysis - the NN parametrization represents a high-order flux discretization (or a high-order correction to flux parametrization) that cannot be applied to non-smooth initial conditions without a flux limiter. In addition, recalling our comparison between the MCL-Lax-Wendroff and MCL-NN parametrizations discussed earlier, the NN parametrization represents flux discretizaiton of the of a higher-order compared to the LW method, since the MCL-NN parametrization reproduces sharp front with high accuracy compared to the MCL-LW simulations with the same resolution. 
\begin{figure}[ht]
\centerline{
\includegraphics[scale=0.25]{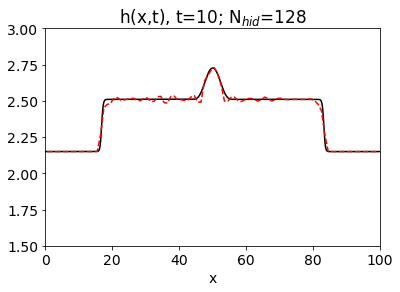}
\includegraphics[scale=0.25]{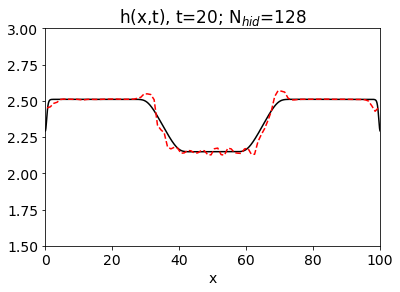}
\includegraphics[scale=0.25]{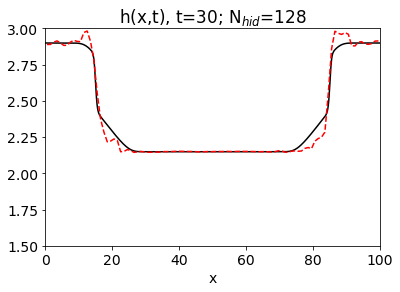}
\includegraphics[scale=0.25]{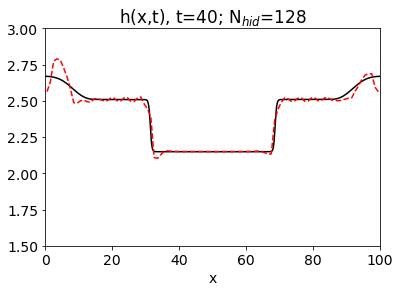}}
\centerline{
\includegraphics[scale=0.25]{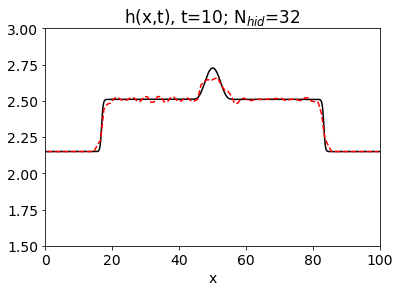}
\includegraphics[scale=0.25]{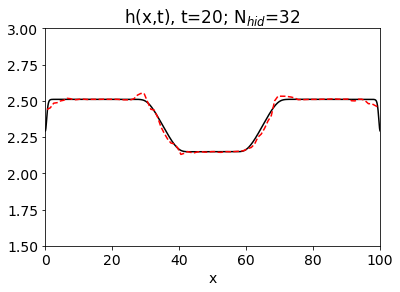}
\includegraphics[scale=0.25]{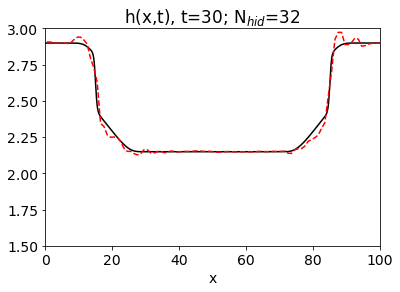}
\includegraphics[scale=0.25]{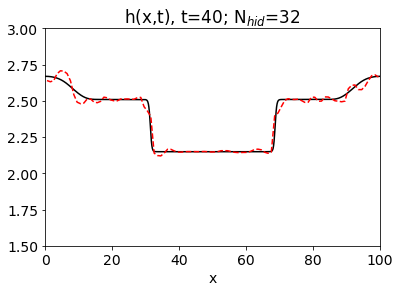}}
\centerline{
\includegraphics[scale=0.25]{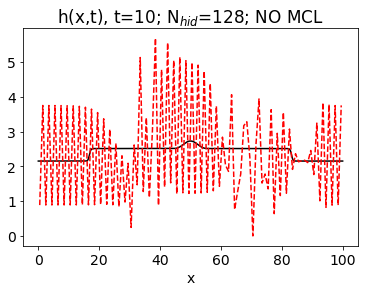}}
\caption{Water height $h(x,t)$ at time $t=10$, 20, 30, 40 in simulations of the full model (black solid line) with $H_0=2.5$ and MCL-NN reduced models with $N_x=100$ ($\Delta x=1$) trained on $H_0=2$ with $n=200$ and $T^{train}=80$ with time-step $\Delta t = 0.05$
(red dashed line). 
Top row - $N_{hid}=128$, middle row - $N_{hid}=32$, respectively. Bottom row - example of $h(x,t)$ at $t=10$ in simulations of the subgrid model with $N_{hid}=128$ without the MCL fix.
Initial conditions with $H_0=2.4$ and $a=0.5$ in \eqref{damic}.
}
\label{fig12}
\end{figure}

}

\section{Conclusions}
\label{sec:conc}
In this work, we developed a neural network reduced model for coarse-grained simulations using the shallow water equations. Starting with a fine-mesh finite-volume discretization, we defined the coarse-mesh variables as ``box'' spatial averages. Then all dynamic variables were decomposed into coarse-scale components (i.e., discretization of the SWE on a coarse mesh) and fine-scale fluctuations. The reduced model uses a relatively simple feed-forward neural network to represent flux corrections due to the subgrid terms (fluctuations).

The neural network is initially trained on a large 
dataset generated using direct numerical simulations with the fine-mesh discretization of the SWE. Our numerical studies indicate that the reduced model is capable of capturing the dynamics of coarse-scale variables for a wide range of initial conditions and for a much longer time compared with the simulation time of each trajectory in the training dataset.
{\mybf
In addition, our results indicate that the NN reduced model can be used outside of the training regime. In particular, we demonstrated good performance of NN subgrid models in regimes where the average water height is varied within $\pm$25\% compared to the training data. }
We would like to point out that the NN parameterization is local in space and requires only a four-point stencil. Local parameterizations have various practical advantages compared to global (e.g., Fourier space) parameterizations of unresolved degrees of freedom. For instance, our approach can be easily generalized to other types of boundary conditions and spatially non-homogeneous problems (e.g., with bottom topography). Moreover, in the spatially non-homogeneous case, neural networks can be trained in parallel using only local four-point stencil data.
We will explore these issues in a subsequent paper for a more realistic ocean flow application.
Finally, our neural network is relatively simple and has only three layers, thus, providing a minimal computational overhead. 
{\mybf Moreover, the network architecture is fairly robust. We tested networks of different sizes and found that smaller networks with 32 neurons in each hidden layer yield very good results. We also demonstrated that the reduced model can be accelerated by using a much larger computational time-step compared to fully resolved simulations.}

The advantage of the approach considered here is that we train a reduced model for subgrid fluxes instead of estimating the whole right-hand side directly. 
Therefore, the reduced NN model is amenable to more traditional numerical techniques, such as flux limiting. Thus, we apply the Monolithic Convex Limiting (MCL) strategy to ensure that the machine learning component of the reduced model does not produce non-physical fluxes. Overall, we believe that combining machine learning with existing tools from numerical analysis has many potential advantages, such as improving the forecasting 
capabilities of ML reduced models, the ability to use simpler neural networks with smaller computational overhead, developing real-time and a posteriori tools for verifying the quality of ML reduced models.

Using numerical simulations, we demonstrated that the flux limiter plays a minimal role when the NN model is well trained {\mybf and utilized in the same regime as the training data (e.g. smooth data).} This is not surprising, since the Neural Network is capable of learning complex functions and, thus, is able to reproduce subgrid fluxes and ensure that they (almost) lie within physical limits. However, when the NN reduced model is not trained sufficiently well, the flux limiter plays a crucial role in suppressing non-physical oscillations. 
{\mybf Thus, the MCL can be used to alleviate the problem of insufficient data and correct NN subgrid parametrizations when they produce non-physical results.}
The computational complexity of the NN- and MCL-NN reduced models is approximately the same since both of them are implemented on a coarse grid. Therefore, a detailed comparison of the numerical results generated by these two models can be used as an indicator of whether the machine-learning model is trained sufficiently well. Such an a posteriori test can then be used to assess the training quality of the neural network and make decisions regarding further training.
{\mybf Moreover, the flux limiter becomes crucial when the NN reduced models are utilized in simulations with with non-smooth initial conditions. This is consistent with the general theory of numerical methods for hyperbolic problems. Since the NN subgrid model aims to represent high-order (in space) flux approximations, flux limiters have to be utilized for non-smooth solutions.}

\section*{Acknowledgments}
I.T. was partially supported by the grant ONR N00014-17-1-2845. The work of A.S. was supported by the German Academic Scholarship Foundation (Studienstiftung des deutschen Volkes). D.K. acknowledges the financial support by the German Research Foundation (DFG) under grant KU 1530/30-1.

We also would like to thank anonymous referees for their valuable suggestions that lead to a considerable improvement of our paper.

\clearpage

\appendix

{\mybf

\section{Running Time}
\label{app:rt}
The additional information provided in this appendix gives a rough idea about the running times of different models. 
The code that we used for SWE simulations 
is implemented in Python.
It was executed on a 56-core Intel(R) Xeon(R) CPU E5-2690 v4 2.60GHz Linux machine. The running times for a single simulation with $T=500$ are presented in Table \ref{tabrt}. These results take CPU utilization into account and are equivalent to a single-CPU run for all models. The wall-time for MCL-NN128 and MCL-NN64
is approximately equivalent to the running times of
MCL-NN32 and MCL-NN16.
Simulations using the MCL-NN128 and MCL-NN64 models are automatically parallelized by PyTorch. This parallelization might introduce substantial additional overhead.
\begin{table}
\centering
\begin{tabular}{c|c|c|c|c}
LLF & MCL-NN128 & MCL-NN64 & MCL-NN32 & MCL-NN16 \\
\hline
493  & 1943  & 1857  & 177  & 165 
\end{tabular}
\caption{Running time (seconds) for a single simulation with $T=500$ of the direct solver (LLF) with $N_f=2000$ points in space and time-step $\Delta t=0.005$ and MCL-NN models with $N_f=100$ points in space and $\Delta t=0.05$. Running times are equivalent to a single-CPU run for all models; wall times should be divided by the corresponding CPU utilization.
PyTorch automatically parallelizes simulations of MCL-NN128 and MCL-NN64 models with CPU utilization 957\% and 991\%, respectively. Simulations of MCL-NN32 and MCL-NN16
are not parallelized.}
\label{tabrt}
\end{table}

Simulations with MCL-NN32 are approximately 2.7 times faster than simulations using the fully-resolved model (LLF). Therefore, a substantial acceleration of the MCL-NN reduced models can be achieved by using a much larger time-step, compared with the direct numerical simulation. This speedup is a direct consequence of using a much coarser computational mesh in simulations with reduced models.

}


\end{document}